\documentclass[preprint,times]{aastex631}

\usepackage[T1]{fontenc}
\usepackage{savesym}
\savesymbol{tablenum}
\usepackage{graphicx}
\usepackage{xcolor}
\usepackage{siunitx}
\usepackage{amsmath}
\usepackage{tabularx}
\usepackage{graphicx}
\usepackage[figuresright]{rotating}
\usepackage{color}  % Used only by the \remark{} macro defined below.
\usepackage{subfiles} % To be able to compile included files separately

\newcommand{\K}{\unit{K}}

\newcommand{\hii}{H{\,\textsc{ii}}}
\newcommand{\oiiRLs}{O{\,\textsc{ii}}}
\newcommand{\oiii}{[O{\,\textsc{iii}}]}
\usepackage{CJKutf8}
\received{\today}

\begin{document}
\begin{CJK*}{UTF8}{gbsn}
\title{Analyzing the Abundance Discrepancy Problem in \hii\ Regions with Photoionization Modeling}

\author[0000-0002-9220-0039]{Ahmad Nemer}
\affiliation{New York University Abu Dhabi, PO Box 129188, Abu Dhabi, UAE}
\affiliation{Center for Astrophysics and Space Science, NYU Abu Dhabi, PO Box 129188, Abu Dhabi, UAE}

\author[0000-0002-6972-6411]{J. E. M\'endez-Delgado}
\affiliation{Instituto de Astronom\'ia, Universidad Nacional Aut\'onoma de M\'exico, Ap. 70-264, 04510 CDMX, M\'exico}
\affiliation{Astronomisches Rechen-Institut, Zentrum f\"ur Astronomie der Universit\"at Heidelberg, M\"onchhofstr.\ 12-14, D-69120 Heidelberg, Germany}

\author[0000-0002-8883-6018]{Natascha Sattler}
\affiliation{Astronomisches Rechen-Institut, Zentrum f\"ur Astronomie der Universit\"at Heidelberg, M\"onchhofstr.\ 12-14, D-69120 Heidelberg, Germany}

\author[0000-0003-4218-3944]{Guillermo A. Blanc}
\affiliation{Observatories of the Carnegie Institution for Science, 813 Santa Barbara Street, Pasadena, CA 91101, USA}
\newcommand{\UChile}{\affiliation{Departamento de Astronom\'{i}a, Universidad de Chile, Camino del Observatorio 1515, Las Condes, Santiago, Chile}} 
\UChile

\author[0009-0000-3962-103X]{Amrita Singh} 
\UChile

\author[0000-0001-6551-3091]{Kathryn Kreckel}
\affiliation{Astronomisches Rechen-Institut, Zentrum f\"ur Astronomie der Universit\"at Heidelberg, M\"onchhofstr.\ 12-14, D-69120 Heidelberg, Germany}

\author[0000-0003-4679-1058]{Joseph D. Gelfand}
\affiliation{New York University Abu Dhabi, PO Box 129188, Abu Dhabi, UAE}
\affiliation{Center for Astrophysics and Space Science, NYU Abu Dhabi, PO Box 129188, Abu Dhabi, UAE}
\affiliation{Center for Cosmology and Particle Physics, New York University, 726 Broadway, room 958, New York, NY 10003}

\author[0000-0002-7339-3170]{Niv Drory}
\affiliation{McDonald Observatory, The University of Texas at Austin, 1 University Station, Austin, TX 78712, USA}

\correspondingauthor{Ahmad Nemer}
\email{abn275@nyu.edu}



\begin{abstract}

Understanding the complex ionization structure and chemical composition of \hii\ regions poses a significant challenge in astrophysics. The abundance discrepancy problem, characterized by inconsistencies between abundances derived from recombination lines (RLs) and collisionally excited lines (CELs), has long been a puzzle in the field. In this theoretical study, we present novel photoionization models that incorporate temperature, density, and chemical inhomogeneities within a single cloud to comprehensively address this discrepancy. By accounting for the intricate interplay between ionization, excitation, and chemistry, our models successfully reproduce both observed RLs and CELs with with an average difference between our models and the observations of 25\% -- within uncertainties inherent in Galactic archival long-slit and new SDSS-V Local Volume Mapper observations. Through comparisons between generic inhomogeneous model predictions and observations, demonstrating the ability of our theoretical framework to analyze the abundance discrepancy problem within \hii\ regions. Our results highlight the importance of incorporating spatially resolved temperature, density, and chemical structures when interpreting the physical processes governing emission line spectra in these astrophysical environments. 
\end{abstract}



\section{Introduction}\label{sec:intro}

The analysis of emission lines in photoionized nebulae serves as a crucial method for understanding the chemical composition of the interstellar medium (ISM) both within our Galaxy \citep{Peimbert:78,Shaver:83,Rudolph:97,Pilyugin:03,Esteban:17,Esteban:22,MendezDelgado:20,Arellano+20}, and especially extragalactic systems \citep{Rayo:82, Vilchez:88, vanZee:98, Berg:13, Kreckel:19, Esteban:20}. \hii\ regions are locations of star formation, where hydrogen is ionized by young massive stars \citep{Osterbrock+2006}. Due to their high surface brightness, these regions serve as tracers of the chemical evolution of the universe as they are visible even at very large cosmological distances \citep{Arellano+22, Topping:24, Sanders:24}. 

Chemical elements heavier than hydrogen and helium, are forged in nucleosynthetic processes related to the life and death of stars, and are subsequently released into the ISM. Therefore, the determination of the abundance of these elements allows us to understand various nucleosynthetic phenomena and quantitatively establish the significance of processes such as the star formation rate and nucleosynthetic yields \citep{Matteucci:86, Carigi:05, Kobayashi:06, Kobayashi:20} . While many ions emit bright emission lines, particularly at optical wavelengths, which can be used to infer the gas-phase chemical abundance, interpretation of these lines in terms of absolute chemical abundances remains challenging due to  the abundance discrepancy (AD) problem.

The AD problem arises when the abundances derived from collisionally excited lines (CELs) are systematically lower than those inferred from metal recombination lines (RLs) \citep{GarciaRojas:07b, SimonDiaz:11, Mendez+2022}. While CELs arise from transitions induced by collisions between electrons and ions, RLs originate from transitions resulting from the recombination of ions with electrons \citep{Osterbrock+2006, Peimbert+2017}. CEL intensities exhibit an exponential dependence on temperature, while metal RLs have a linear dependence that cancels when estimating abundances by comparing them with hydrogen RLs. This has led to one of the most popular explanations for the AD problem being the presence of temperature fluctuations, which would affect mainly the CELs, as proposed by \citet{Peimbert+1967}. This paradigm suggests that abundance determinations using RLs should be more accurate than those relying on CELs, which would be systematically underestimated. However, their faintness limits RL-based abundance measurements to a small number of ionized nebulae. Therefore, understanding the thermal and density structure of ionized nebulae holds significant importance achieve accurate determinations of chemical abundances in the universe.

Other hypotheses have been also suggested to explain the AD problem, such as erroneous atomic data \citep{Nemer+2019}, a non Maxwell-Boltzmann distribution of free electrons \citep{Nicholls+2012}, or chemical variations \citep{Torres-Peimbert+1980, Liu+2000}. The presence of metal-rich (i.e., hydrogen-poor) inclusions is another popular scenario to explain the AD problem, specially in the case of planetary nebulae \citep{GarciaRojas:22,GomezLlanos:24}. According to this hypothesis, the existence of metal-rich inclusions within the photoionized gas enhances the emission of metal RLs due to strong thermal cooling, while CELs are emitted outside these metal-rich pockets, in gas of ``normal'' metallicity. Several authors have developed two-phase photoionization models that successfully replicate both RLs and CELs emissions in \hii\ regions \citep{Tsamis+2005}. However, the source of such metal-rich inclusions has not yet been firmly established.

Despite their great capacity and usefulness, simple photoionization models often fail to capture the complex internal structures of \hii\ regions adequately \citep{Jin+2023,Marconi+2024}. The typical assumption of constant density conditions, as well as simple geometries, may not be realistic in many cases \citep{Filippenko+1985,Osterbrock+1989}. The photo-ionization models solve the thermal equilibrium equations and calculate the temperature at each position so the temperature is typically uniform and changing smoothly which results from the radiative transfer in a homogeneous medium. For instance, \citet{Jin+2023} made simple photoionization models to compare with spatially resolved observation of \hii\ regions and concluded that more sophisticated treatment of the plasma conditions is required to explain the observed complex internal gradients in electron temperature and density. From the observational side, by comparing \oiiRLs-RLs and \oiii-CELs \citet{MendezDelgado:23b} found that the effect of temperature inhomogeneities on Galactic and extragalactic \hii\ regions may be significant and provided evidence that it could explain the AD problem. The existence of such inhomogenieties could be due to shocks or stellar feedback from massive stars.

Considering the potential importance of the temperature, density or chemical inhomogeneities in real nebulae, in this paper we present a comprehensive analysis of the AD problem in \hii\ regions, incorporating state-of-the-art observations and theoretical models. We explore the impact of large scale temperature, chemical and density inhomogeneities on the derived abundances and emission line fluxes. We compare our 1D photoionization models with optical observations of galactic \hii\ regions with relatively simple morphologies obtained with single slit instruments to show good agreement in reproducing both the observed RLs and CELs. Finally, we make comparisons with spatially resolved optical spectra for the galactic \hii\ region M\,20 to show a good match in the overall structure of the nebula and emission characteristics. 

\section{Observations}

In this work, we will focus on the Galactic regions M\,17, M\,20, and NGC\,2579. These regions are located at Galactocentric distances of around 6.5 kpc, 6.8 kpc, and 10.8 kpc, respectively \citep{Mendez+2022}. This suggests that the first two regions are likely to have slightly super-solar metallicities, given the metallicity gradient of the Milky Way \citep{Arellano+20,Mendez+2022}, while the last region is expected to be slightly sub-solar. M\,17 and M\,20 are thought to have similar chemical compositions considering their Galactocentric positions; however, their ionization states are very different. M\,17 is a high-ionization nebula \citep{Garcia+2007} (defined by the abundance ratio O$^{2+}$/(O$^{+}$ + O$^{2+}$)), whereas M\,20 is a low-ionization nebula \citep{Garcia+2006}. This allows us to study the impact of density structure and the ionizing source (key parameters in defining the ionization parameter of the gas) on photoionization models. The selected regions have relatively simple geometries, with M\,20 being the most notable case due to its nearly spherical shape. 

For this study, we have analyzed archival deep echelle observations from UVES at the Very Large Telescope, as integrated, dereddened and reported by \citet{Garcia+2007}, \citet{Garcia+2006}, and \citet{Esteban+2013} for M\,17, M\,20, and NGC~2579, respectively. Although these observations are spatially limited, they are deep enough to detect the ultra-faint \oiiRLs-RLs necessary for studying the AD problem. 

In the case of M\,20, we include spatially resolved observations obtained as part of the Sloan Digital Sky Survey V (SDSS-V), Local Volume Mapper (LVM) project \citep{Drory:24}. These observations were taken during September 27 to 30, 2023 % September 27, 29 and 30)
in 15-minute exposures and were reduced using the standard routines described in \citet{Drory:24}.These observations consist of eight combined 15-minute exposures centered at 
$\text{RA} = 18^{\text{h}}02^{\text{m}}32.16^{\text{s}}, \quad \text{DEC} = -22^\circ58'52.96''$, covering the entirety of M\,20 within a half-degree field of view using 1801 fibers of 35 arcseconds in diameter. This setup yields R4000 spectra across the full optical wavelength range (3600–9800 \AA). The data are processed using the LVM Data Reduction Pipeline (DRP), which calibrates each spectrum for wavelength and flux, resulting in fully calibrated row-stacked spectra. Although our sky subtraction algorithm is still under development, any potential sky contamination effects in the lines used for this study are negligible. The physical properties (such as electron density, temperature and oxygen abundance) were calculated from the Gaussian fitted and reddening corrected emission lines using the \textsc{pyneb} \citep{Luridiana+2015} \textsc{python} package. More details and a complete analysis of the optical spectrum will be presented in Sattler et al. (in prep.), while in this work we will focus on photoionization modeling with the aim of understanding the AD problem.

\section{Methods}\label{sec:methods}

\subsection{Photoionization Models}

We perform photoionization modelling using \textsc{cloudy} \citep{Ferland+2017} to make predictions of the optical emission line spectrum for nearby Galactic \hii\ regions. \textsc{cloudy} is a photoionization code that is used to
solve the radiative transfer, thermal equilibrium, ionization balance, and statistical equilibrium equations to calculate the ionization structures of the ISM plasma and the level populations responsible for the line emission. Th 1D code generates the nebular radial variation of the electron density, temperature, ionic fraction, and emissions of 30 elements across the nebula. We adopt atomic data from version v8.0 of the \textsc{chianti} database \citep{DelZanna+2015}. We can also control the number of energy levels included for the relevant atoms and ions  to account for all the atomic processes responsible for the line formation. We include all available energy levels (usually hundreds) for the species responsible for the lines commonly used in abundance determinations; namely carbon, oxygen, nitrogen, and sulphur in addition to hydrogen and helium,. 

\textsc{cloudy} propagates both the incident radiation field (specified by the user), and the diffuse radiation field that results from interaction of the gas with the incident light. We choose our incident radiation field from the Atlas stellar atmospheres models \citep{Castelli+2005}, which are plane-parallel, hydrostatic, and in Local Thermal Equilibrium (LTE). They utilize the latest opacity distribution functions. These grids encompass various metallicities, spanning from log(Z) = +0.5 dex to -2.5 dex, all with a turbulent velocity of 2~km~s$^{-1}$. 

We employ spherical geometry models for our comparisons. \textsc{cloudy} is specifically designed to produce both plane-parallel and spherical nebular models. Plane-parallel models are typically suited for nebulae with a radius much larger than their thickness (such as the Orion Nebula, \citealt{Ferland:01}), while spherical models are better suited for clouds powered by a point source at their center, such as the \hii\ regions considered in this work. Since these are expected to be at roughly solar metallicity, we simulate with stellar atmosphere models of solar metallicity; i.e. log(Z) = 0. We refer the reader to \citet{Castelli+2005} for more information about the details of the solar metal abundances used in those models. Since there is a metallicity gradient towards the Milky Way galactic center, we expect nebulae closer to the center than the sun (e.g., M\,20) to have slightly super solar abundances but the effect should be negligible. 

\begin{figure}[htp]
\centering
%\hspace*{-1in}
\includegraphics[width=\textwidth]{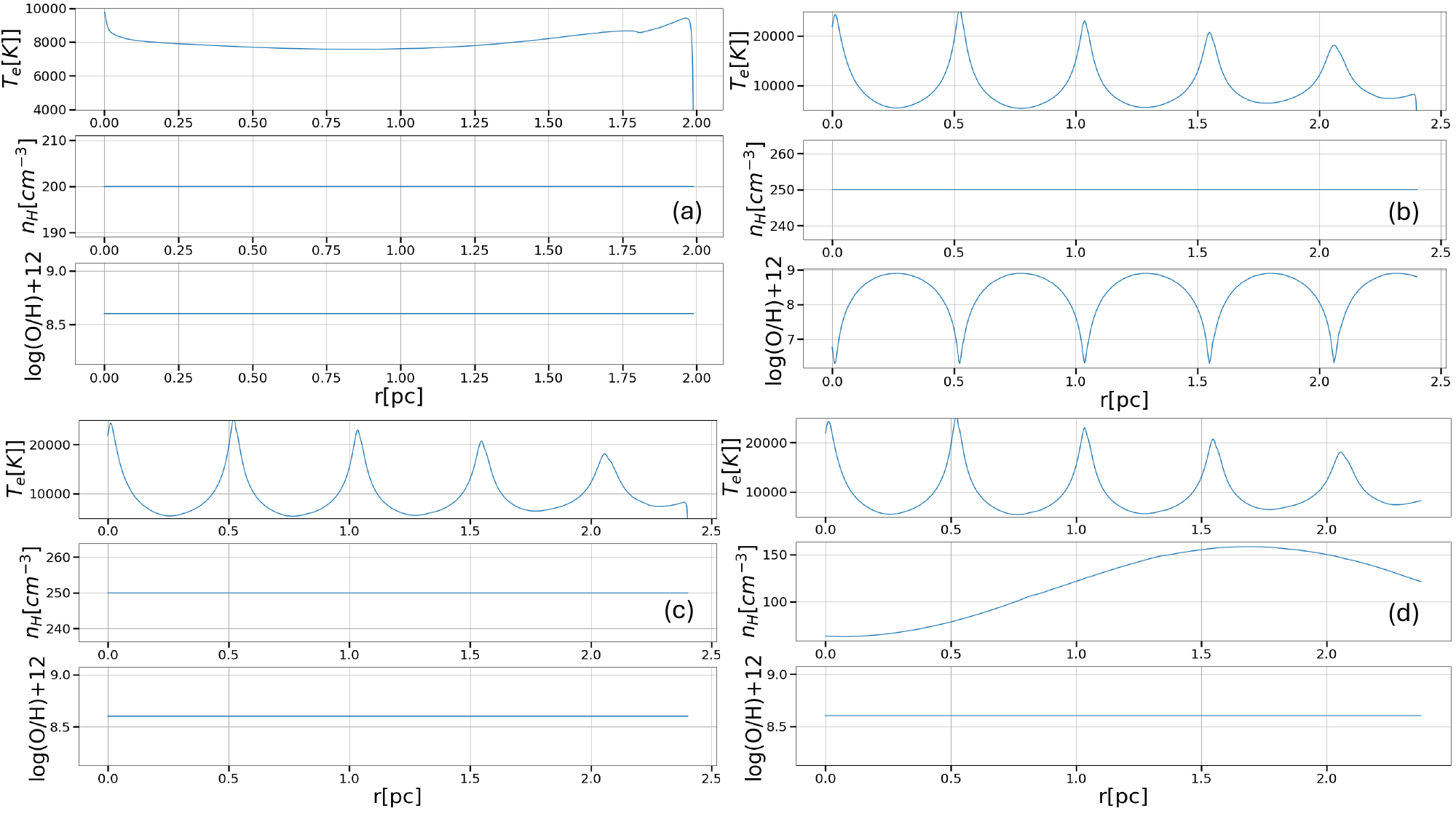}
\caption{Temperature, density, and chemical abundance profiles of a standard HII region homogeneous model (a), with variations where we introduce chemical (i.e. family 1)(b), temperature (family 2)(c), and temperature and density inhomogeneities (family 3)(d).}
\label{fig:1}
\end{figure}

In Fig.~\ref{fig:1} we show a homogeneous model (panel a), and then introduce inhomogeneities to show the effect that they have on the general properties of the simulation. For one family of inhomogeneous models, hereafter referred to as family (1), we adjust the gas abundances by introducing small and large-scale variations, scaling the fiducial homogeneous model with a radial-dependent factor. (Fig.~\ref{fig:1}, panel b). The dimensionless scaling factor follows an asymmetrical sinusoidal form, characterized by adjustable peak, trough and period parameters that account for both small- and large-scale variations across the nebula. The peak values are specified to range between 0.1 and 10, and trough values between 0.001 and 1.

Moreover, the temperature can be taken as an input in \textsc{cloudy}, as opposed to being solved for, so that the emission calculations are set by the specified temperature in each zone. Hence, for a second family of models, family (2), we introduce the shape of the temperature profile that would best reproduce the observed spectrum (Fig.~\ref{fig:1}, panel c). We obtain a semi-realistic temperature profile from a preliminary \textsc{cloudy} run, where we apply a sinusoidal scaling factor to the abundances, similar to the approach used in family (1). This resulting temperature profile is then used as input for the main \textsc{cloudy} run in family (2), allowing the temperature profile to be parameterized by the abundance scaling factor, without altering the abundances themselves. Lastly, we have a third family of models, family (3) where we specify the shape of the temperature profile, as in family (2), and also introduce a sinusoidal density profile (Fig.~\ref{fig:1}, panel d). This approach becomes clearer when comparing panels b, c, and d in Fig.~\ref{fig:1}. Panel b represents a family (1) run in which the temperature changes as a result of change in abundances, panel c is for a family (2) run where we use the temperature output of panel b as input without changing the abundances, and panel (d), representing family (3), is the same as panel c but with a sinusoidal density profile.

 For all the models in this work we use Orion \hii\ region abundances (as implemented by \textsc{cloudy}) which is the average of several abundance studies for the region \citep{Baldwin+1991,Rubin+1991,Osterbrock+1992}. We only show the oxygen abundance since the scaling factor applies uniformly to the entire abundance set; i.e. oxygen, nitrogen, and carbon will be multiplied by the same scaling factor. This is a crude representation of high metallicity regions since various processes affect different elemental abundances differently, hence they are not expected to be enhanced by the same factor. Specifically tailored models should have a scaling factor for each element separately. We choose a source with effective temperature $T_{eff}$ between 30,000~K and 50,000~K with the luminosity log($L^*$) in the range 38 - 42 as appropriate for \hii\ regions \citep[and references therein]{Peimbert+2017}. We included grains in the simulation as they would affect both the thermal solution and the line fluxes due to scattering and absorption effects. We include Orion grains (as implemented by \textsc{cloudy}) and use a scaling factor between 0.1 - 5 since the dust content is not constrained well from observations. It is notable that the observations we compare with are all de-reddened so we do not have to worry about extinction effects.\\ 

\subsection{Search Process}

We first construct a grid of models using the above mentioned simulation parameters; namely the luminosity and temperature of the source, the scaling of the grain content, the abundance set, the homogeneous density parameter (around which inhomogeneities are injected), and the parameters of the sinusoidal scaling parameters for the density and/or the abundances depending on the family of models being simulated. We use the Latin hypercube sampling method \citep{Jared+2012} to generate semi-random samples of those parameters withing the above mentioned ranges. The grid of input parameters are coupled with their matching spectra to create a larger grid of input/output data. The predicted spectrum consists of a list of wavelengths and their corresponding integrated fluxes for the 17 lines listed in Tab.~\ref{tab:1}. These lines include a range of CELs and RLs, whose ratios serve as important diagnostics for temperature or density in different components of the \hii\ region. Additionally, some RLs are included that are necessary for determining the abundances of their respective elements.

Specifically, the density-sensitive line ratios are [O~\textsc{ii}] $\lambda3729/\lambda3726$ and [S~\textsc{ii}] $\lambda6716/\lambda6730$. The temperature-sensitive ratios are [O~\textsc{iii}] $\lambda4363/(\lambda4959+\lambda5007)$, [N~\textsc{ii}] $\lambda5755/\lambda6584$, and He~\textsc{i} $\lambda7281/\lambda6678$. The ratios sensitive to both temperature and density are [S~\textsc{ii}] $(\lambda\lambda 4069+4076)/(\lambda\lambda 6716+6730)$ and [O~\textsc{ii}] $(\lambda\lambda 7320+7330)/(\lambda\lambda 3726+3729)$.
All the lines fluxes are normalized to $H\beta$ mimicking the observations we compare with. Finally, we identify the best-fit model of the de-reddened integrated emission-line spectrum with the minimal
$\chi^2$ value, where $\chi^2$ is defined as:
\begin{equation}
    \chi^2 = \sqrt{\sum_\lambda((F_\lambda^{obs} - F_\lambda^{sim})/F_\lambda^{obs})^2}
\end{equation}
%$\chi^2 = \sqrt{\sum_\lambda(F_\lambda^{obs} - F_\lambda^{sim}/F_\lambda^{obs})^2}$
where $F_\lambda^{obs}$ and $F_\lambda^{sim}$ are the observed and simulated integrated fluxes of the line at $\lambda$.

%\newpage

\section{Results}\label{section:results}

% Here we compare the predictions of our models with high quality observations of galactic \hii\ regions. Traditionally, these observations are made with a single slit/fiber instruments, so they focus on a defined section (sometimes the brightest) of the ionized gas. Recently, there has been rapid development in the instruments capable of spatially resolved observations of nearby photoionized nebulae such as Multi Unit Spectroscopic Explorer (MUSE) and the Local Volume Mapper (LVM) part of the Sloan Digital Sky Survey (SDSS-V). Spatially resolved observations present a better opportunity to study the spatial variations of nebular conditions and allows us to impose better constraints on our models. \\

\subsection{Single slit observations}
In the ideal picture of a Strömgren sphere around a single central source, any variations in the physical conditions across the observed section should represent the variations across the spherically symmetric nebula. In reality, most \hii\ regions have irregular shapes and multiple ionization sources which would lead to  asymmetric variations depending on the viewing angle, slits' sizes, and their location with respect to the central source and the Stromgren radius. Since our models are 1D in nature, we want to avoid variations in nebular conditions across the slit, so our selected systems are relatively simple and spherically symmetric with a single dominant central ionizing source. In each slit we integrate along the line of sight, so we are observing a luminosity weighted average of gas parcels at different 3D radii from the center of the nebula; which 3D radii get sampled depends on how close to the center or to the edge of the nebula you are. When integrating the flux using the 1D models for such nebulae, any geometrical factor is omitted since we normalize to $H\beta$ which applies only if the conditions are homogeneous across the aperture. Namely, we compare our models with single slit observations of the \hii\ regions M\,20 \citep{Garcia+2006}, M\,17 \citep{Garcia+2007}, and NGC\,2579 \citep{Esteban+2013}. \\

The three nebulae cover a range of central stellar temperatures of 39,000~K < $T_{eff}$ < 46,000~K \citep{Garcia+2006, Garcia+2007,Esteban+2013} while the luminosities of the sources range between 38.6 < log $L^*(erg/s)$ < 41.4. The electron temperature of the nebulae derived from the best-fit model range between $4,000 K < T_e < 9,500 K$ and are consistent with the volume-weighted average temperature of the temperature diagnostic line ratios as reported by \citet{Garcia+2006,Garcia+2007,Esteban+2013}. On the other hand, the hydrogen density of the nebula derived from the best-fit model ranges between 150~$cm^{-3}$ < n$_H$ < 1500~$cm^{-3}$; we note that these ranges are rather wide which indicates significant differences between these regions. There is insufficient information about the $T_{eff}$ and $L^*$ in the literature, so it is difficult to constrain these parameters in the models. Moreover, M\,20, for example, has multiple B stars supplying ionizing photons, in addition to the dominant O star. These considerations can be dealt with in a more elaborate model for specific regions; possibly Orion or Rosette which we plan to work on next.

We present common temperature and density line diagnostics to better compare their ratios with the observations. We use \textsc{pyneb} 1.1.13 \citep{Luridiana+2015} as described in \citet{Mendez+2022} to estimate the electron density $(n_e)$ and temperature $(T_e)$ from the same emission lines. In most cases, there is general agreement between the model and the data, but in the case of NGC\,2579 the density and temperature are slightly under-predicted. All the relevant model parameters for each system are reported below the figure describing it. We describe in each figure which family of models was best able to reproduce the observed spectrum. \\

We select the fiducial model based on the key emission lines discussed in the previous section, and their percent mismatches ($\Delta_f$\%) with the observations are reported on Tables.~\ref{tab:1} and \ref{tab:2}. We note that in this simple fitting procedure we select the model with the absolute minimum $\chi^2$ score across all the families, and we don't make any attempts to evaluate or score models with similar $\chi^2$ scores since this paper is about introducing a new method rather than finding tailored models. The average error in the emission flux predictions around $25\%$ with some lines experiencing higher differences while the line ratios typically yield an error $\leq\%10$. These differences are acceptable in accordance with previous theoretical modeling work \citep{Tsamis+2005,Jin+2023,Marconi+2024}. Our models reproduce the observed diagnostic emission lines with high fidelity compared to the homogeneous model, and this comparison ($\Delta_h$\%) is reported in Tables.~\ref{tab:1} and \ref{tab:2}.%,tab:2}.

For M\,20, family (1) lead to higher or lower temperatures (for diminished or enhanced metallicities, respectively) than observed, and family (2) lacks the density gradient we observe. In the case of M\,17, families (2) and (3) required a density profile much higher than observed to explain the emission line spectrum with a similar temperature profile as in family (1). This is an indication that M\,17 requires enhanced metallicities within the emitting region to produce the observed spectrum. Finally, NGC\,2579 was best fit with family (2) of models, because for similar densities to what is observed, the high/lower metallicities produce lower or higher temperatures, and family (3) gave rise to significantly higher temperatures and lower densities to match the observed spectrum.

\begin{figure}[!ht]
    \centering
    \includegraphics[width=.49\textwidth]{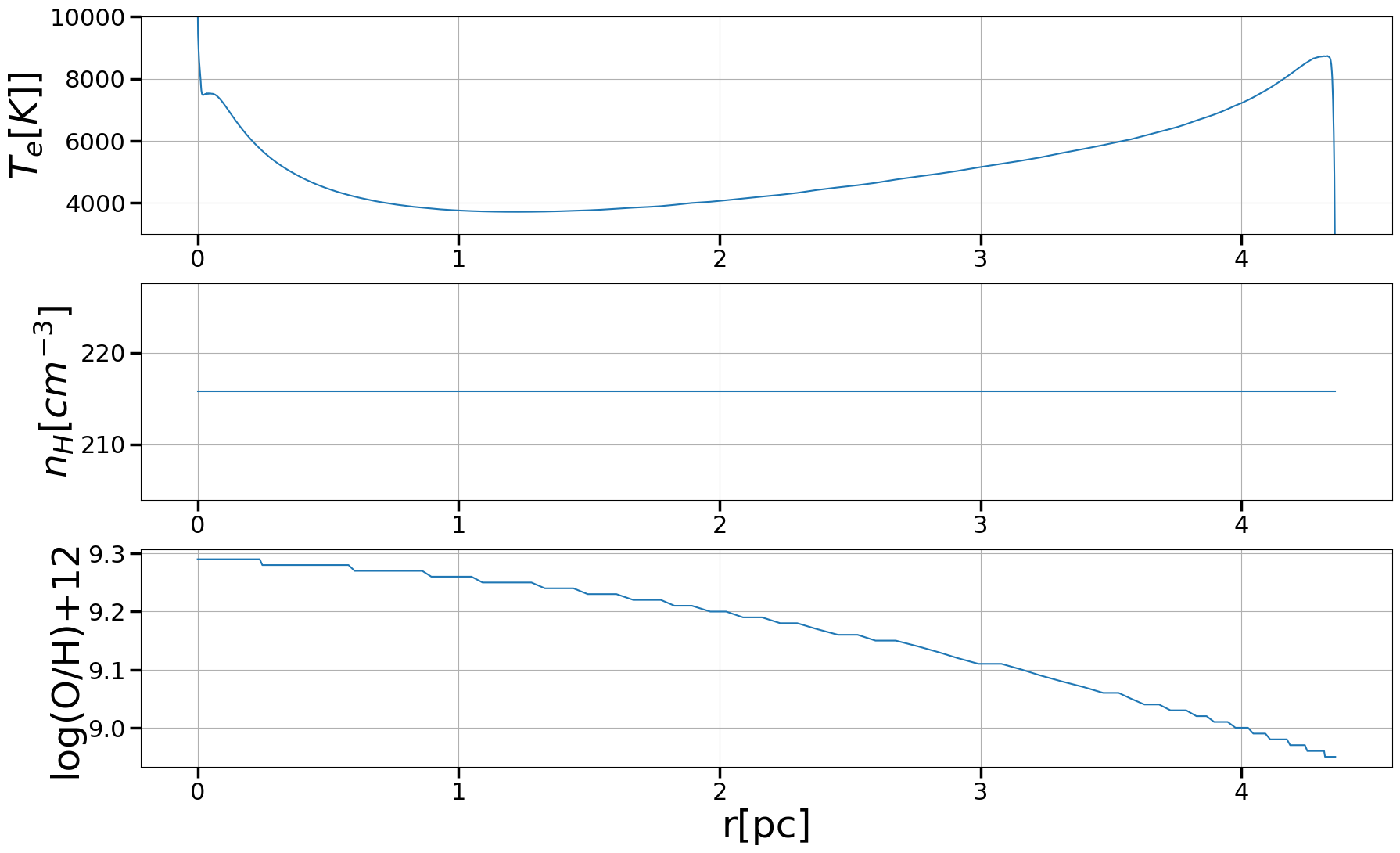}
    \hspace{0cm}
    \includegraphics[width=.49\textwidth]{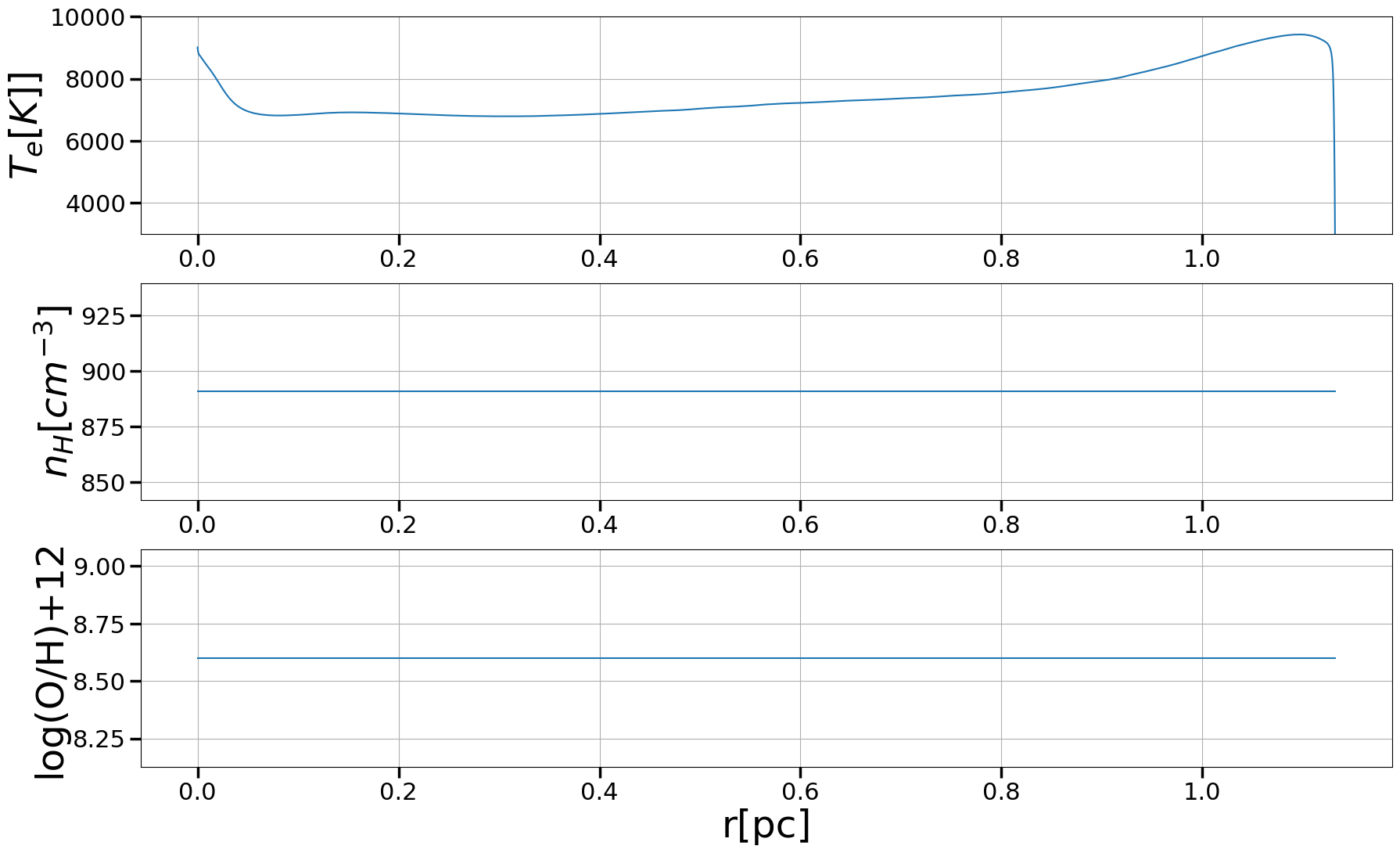}
    \caption{Left: Best fit model for the single slit observation
of M\,17 belonging to family (1) where we introduce chemical inhomogeneities. Right: Best fit model for the single slit observation of NGC\,2579 belonging to family (2) where we introduce temperature inhomogeneities. For both, we show the temperature on top, density in the middle and the chemical abundance profile at the bottom panel.}
  \label{fig:3}
\end{figure}

In the case of NGC\,2579, the [N\,\textsc{ii}] $\lambda5755$ line is not detected. This complicates the analysis since this line is needed to estimate the temperature in the lower ionization zones which could explain the model under-prediction of the physical parameters in this particular nebula. We show (Fig.~\ref{fig:3}) the electron temperature, electron density, and oxygen abundance radial profiles to get a picture on the variations required to simulate these environments. We note that our best-fit model for M\,17 involves an abnormally high oxygen abundance, which is probably unrealistic. This model is intended to simulate the presence of hypothetical high-metallicity clumps that are below the resolution of both the observations and the model. As suggested by \citet{Liu+2000}, these clumps could dominate the emission of heavy element RLs, making such a high-metallicity model the best available candidate.

\subsection{Spatially resolved observations with the SDSS-V LVM}

M\,20 was included in the early science program at the LVM, so we obtained spatially resolved optical spectra for key diagnostic lines. In general, emission lines used for the electron density and temperature diagnostics in our \textsc{cloudy} models, such as [O\,\textsc{ii}]~$\lambda\lambda$3726,29; [S\,\textsc{ii}]~$\lambda\lambda$6716,30; [N\,\textsc{ii}]~$\lambda\lambda$6584,5755 are also detected with a Signal-to-Noise Ratio (SNR) > 5 in M\,20 for several LVM fibers inside its surrounding Strömgren Sphere (Sattler et al. in prep.). The lines of [S\,\textsc{ii}]~$\lambda$4076, [O\,\textsc{iii}]~$\lambda$4363 and He\,\textsc{i}~$\lambda$7281 are either not measured in M\,20 with a good SNR or blended with sky lines.
The lines of C\,\textsc{ii}~$\lambda$4267 and O\,\textsc{ii}~$\lambda$4641 are currently not detected within the LVM data, however with planned future developments of the corresponding data reduction pipeline they may be possible in the future. This pipeline is still undergoing several updates and improvements, and there may be some uncertainties in the early science LVM data related to flux calibration and sky subtraction. However, any potential uncertainties in the flux calibration, which could mainly affect the blue arm, do not impact the determination of the electron density from [O\,\textsc{ii}] 3727/3729. Moreover, they do not alter the overall distribution of temperatures and ionic abundances based on [O\,\textsc{ii}] 3727, 3729, although there could be some offset in these latter two cases without changing the overall distribution. Possible effects of sky subtraction are not expected to be significant for the lines analyzed here.

\begin{figure}[!ht]
    \centering
    \includegraphics[width=.49\textwidth]{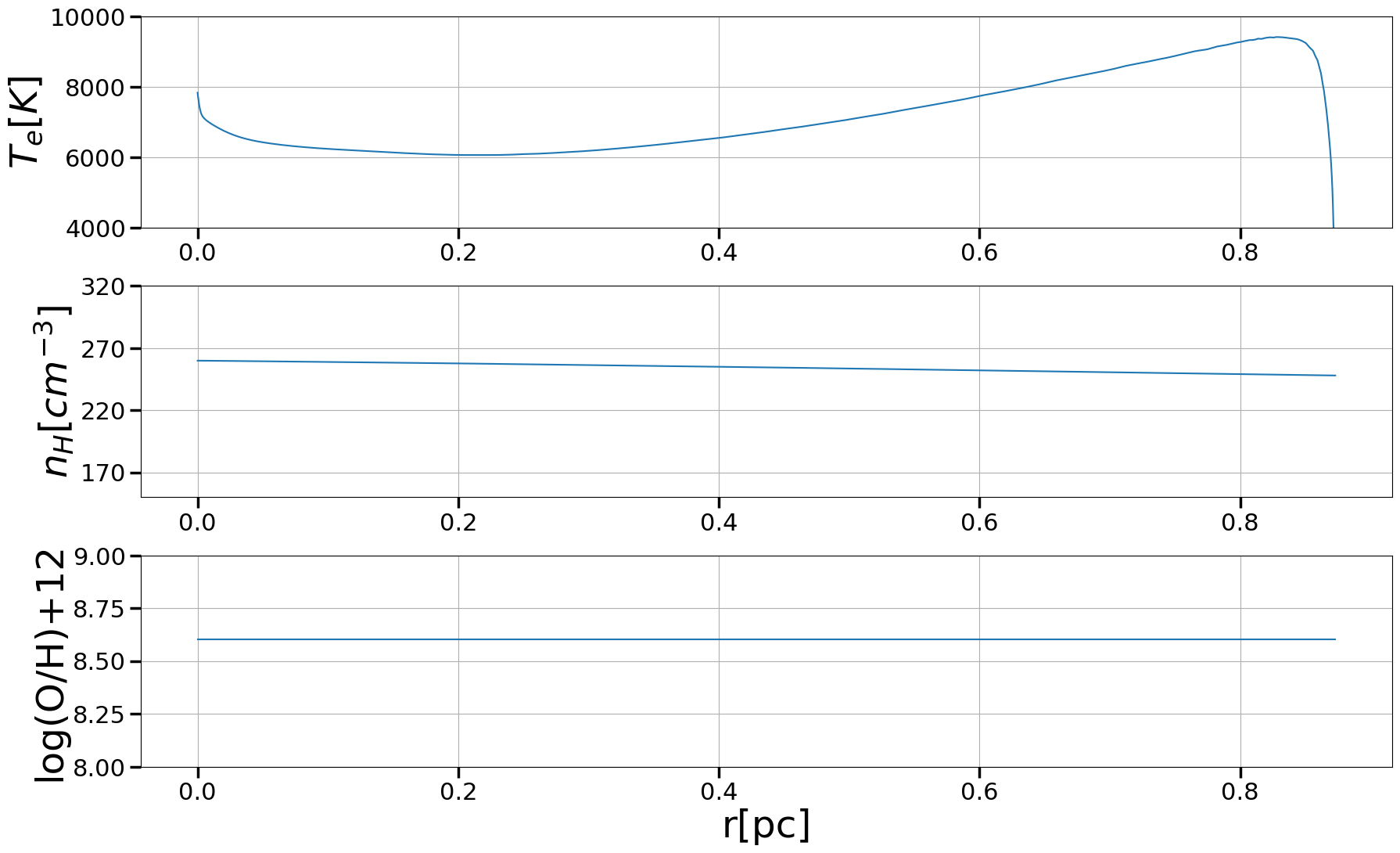}
    \hspace{0cm}
    \includegraphics[width=.49\textwidth]{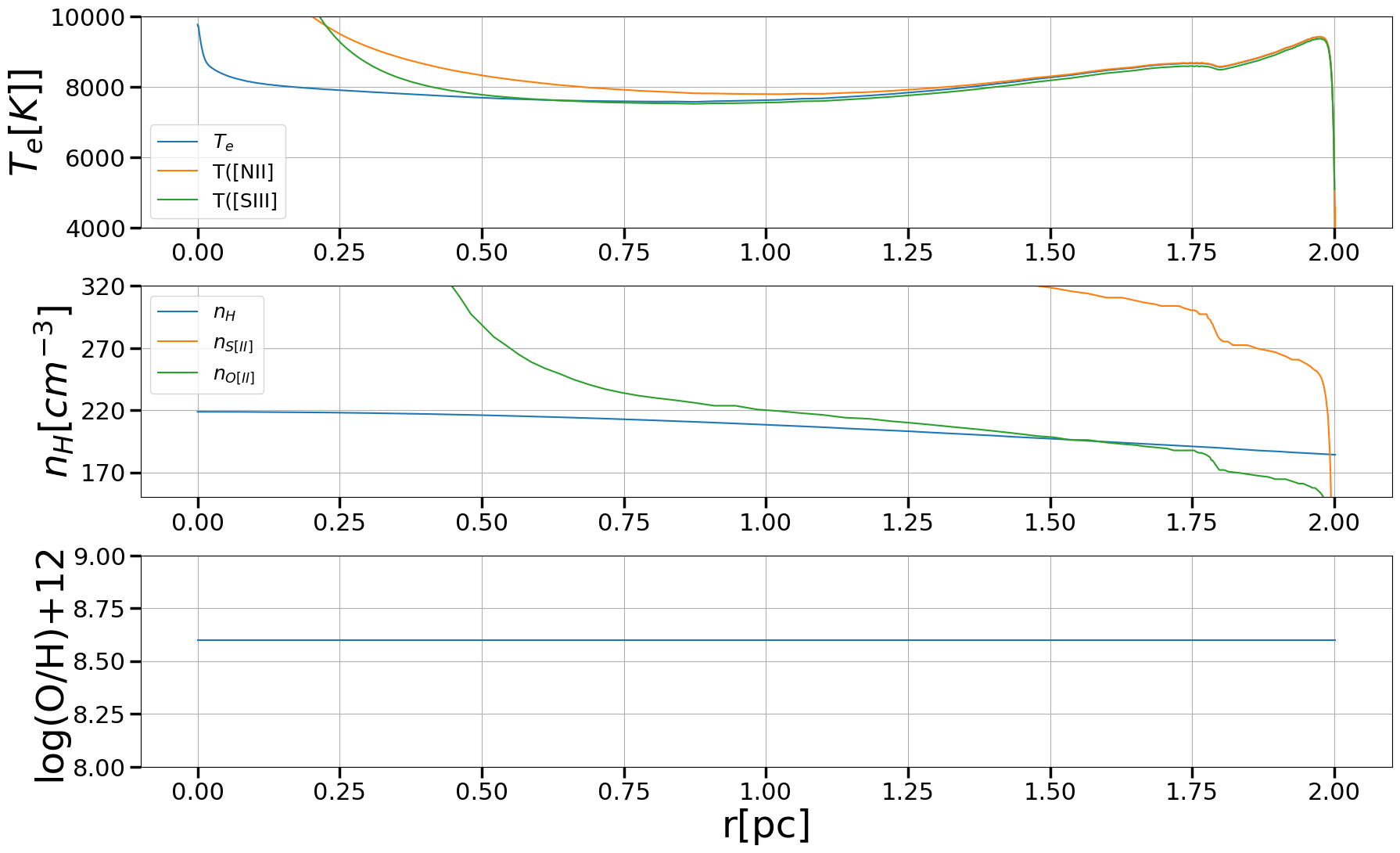}
    \caption{Left: Best fit model for the single slit observation
    of M\,20 belonging to family (3) where we introduce temperature and density inhomogeneities. Right: Best fit model for the single slit observation of M\,20 while considering further constraints from LVM. The density radial profile based on [S\,\textsc{ii}]~$\lambda\lambda$6716,30 and [O\,\textsc{ii}]~$\lambda\lambda$3726,29 doublet ratios, and the temperature based on [N\,\textsc{ii}]~$\lambda\lambda$6584,5755 is shown for comparison with the LVM data For both, we show the temperature on top, density in the middle and the chemical abundance profile at the bottom panel.}
  \label{fig:2}
\end{figure}

\begin{figure}[htp]
\centering
%\hspace*{-1in}
\includegraphics[trim={0in 0.3in 0 0},clip,width=0.49\textwidth,height=3.02in]{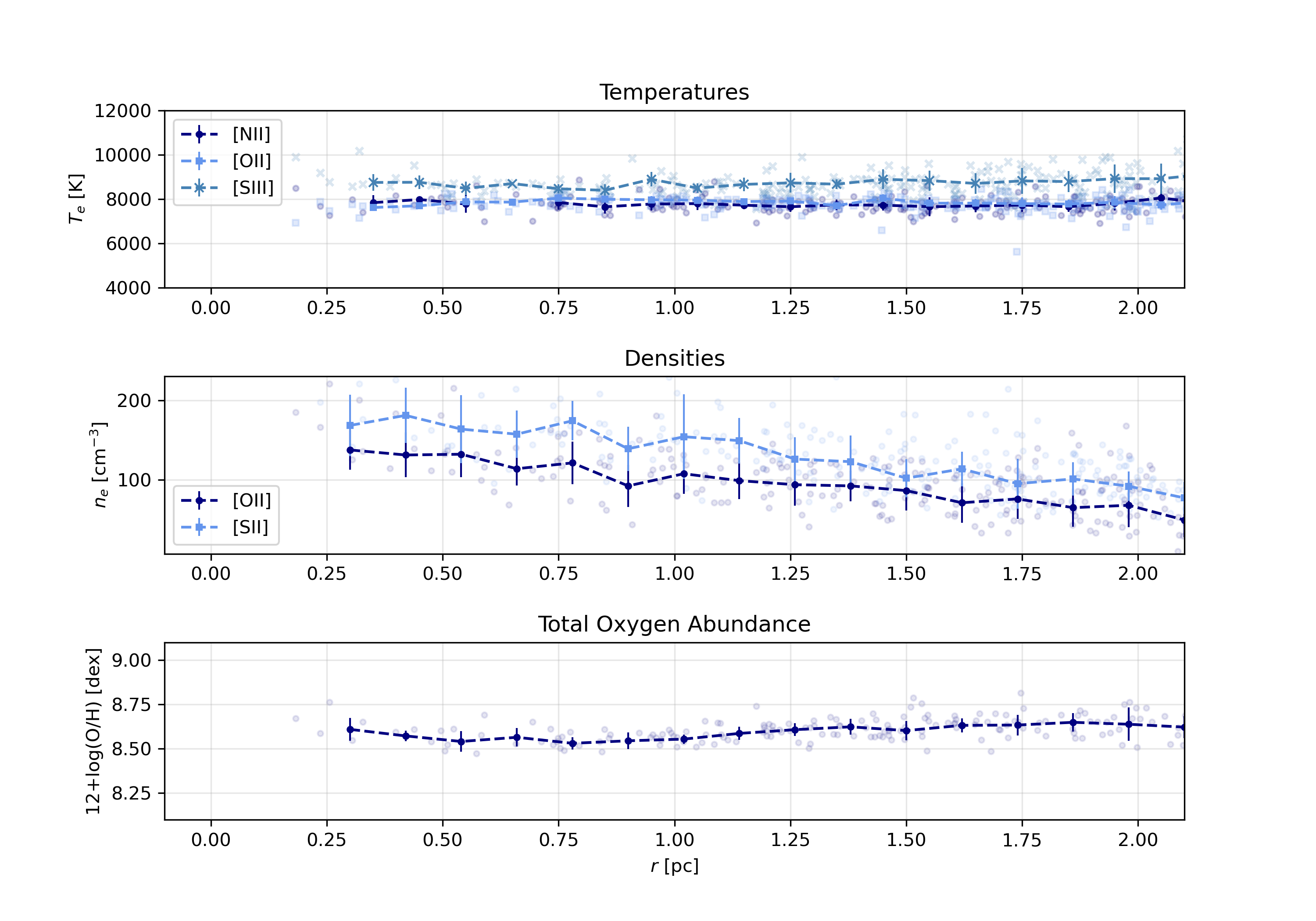}
    \hspace{0cm}
    \includegraphics[width=.46\textwidth,height=2.65in]{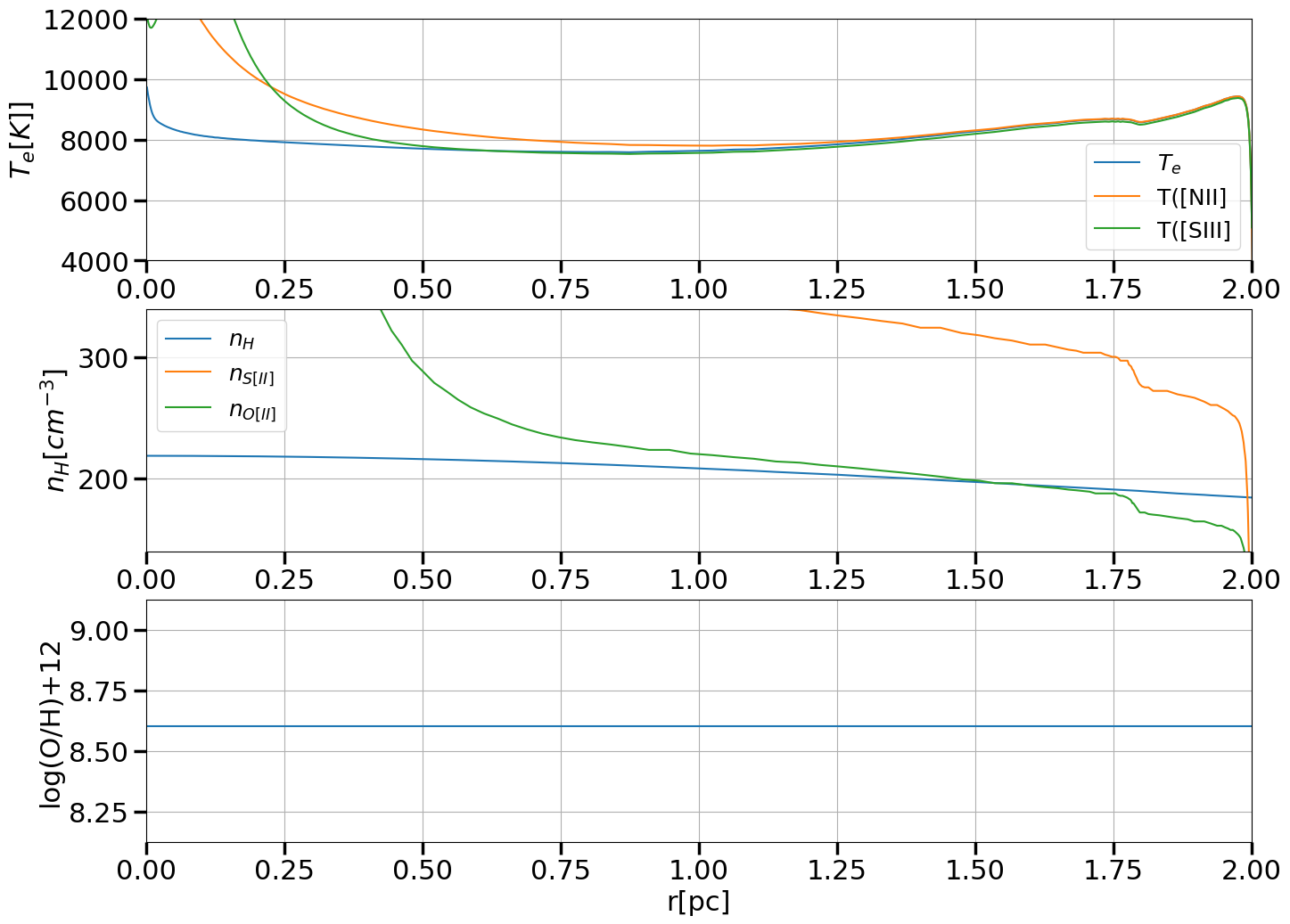}
\caption{Left: Temperature, density, and oxygen abundance profile of M\,20 as detected by the LVM. The electron densities were measured using the [O\,\textsc{ii}] 3727/3729 and [S\,\textsc{ii}] 6717/6731 doublets. To calculate the electron temperatures, we used the resolved mean value of both densities together with the [N\,\textsc{ii}] 5755/6584, [O\,\textsc{ii}] (3727+3729)/7330 and [S\,\textsc{iii}] 9531/6312 lines. The final total oxygen abundance was then inferred with the mean resolved density and the mean temperature of [N\,\textsc{ii}] and [O\,\textsc{ii}] for the single ionized oxygen and [S\,\textsc{iii}] temperature for the double ionized oxygen. Right: Best fit model for the single slit observation of M\,20 while considering further constraints from LVM; the same as Fig.~\ref{fig:2}.}
\label{fig:4}
\end{figure}

Figs.~\ref{fig:2}, \ref{fig:4} compare the LVM observations with the best-fit model for this data. We do not fit the radial profile of the line fluxes from CLOUDY to the observed LVM radial profile. Instead we compare the observed LVM profile to the best-fit to the single-slit spectra while further constraining the ranges of the temperature and density diagnostic lines by the LVM as well as the radial extent of the photoionized gas to match the LVM data.

Tab.~\ref{tab:1} reports the comparison between the updated best-fit model integrated line fluxes and the single slit integrated fluxes while the radial profiles are presented in Fig.~\ref{fig:2}. There are obvious discrepancies between the best-fit model based on the \citet{Garcia+2006} observations alone and the one further constrained by the LVM data, but the general properties are the same. While the density estimates for the former are somewhat higher, the negative gradient is a feature of both models. It is also worth noting that the [S\,\textsc{ii}] density diagnostic has a higher estimate than the [O\,\textsc{ii}] line ratio in the updated best fit model. The oxygen abundance is fairly uniform and the temperature is constant around 8,000~K with a slight increase towards the outer edge, but the line fluxes of the key diagnostic lines agree well with the \citet{Garcia+2006} data. One complication to this comparison is that the [O\,\textsc{ii}] $\lambda$4363 diagnostic line was detected by \citet{Garcia+2006} and not by the LVM, and this line is important to probe the temperature in the high ionization zone close to the source.

\section{Discussion and Conclusion}\label{section:discussion}

The AD problem is a long-standing predicament that has been hampering the accurate estimation of chemical abundances in photoionized plasmas. This problem arises when the temperature—and more importantly, the chemical abundances—inferred by the CELs are different from those inferred by the metal RLs. Many reasons have been suggested to explain this phenomenon, but the most prevailing argument is the presence of inhomogeneous physical conditions within the gas, a scenario supported by recent observational evidence \citep{Mendez+2022,Jin+2023}. Although the inhomogeneous gas scenario has gained popularity in recent years, theoretical work is lagging behind, and there is little recent work to address these new observational constraints.

In this paper, we introduce a new theoretical framework to model H\,\textsc{ii} region spectra, focusing on studying the sensitivity of the predicted spectra to model parameters. Specifically, we present \textbf{the first theoretical models that include different types and scales of inhomogeneities within a single cloud model} using \textsc{cloudy}. This is a first attempt to construct such models and explore how they can be used and improved, without aiming to tailor exact models to the studied regions.

In this proof-of-concept work, we select a few simple H\,\textsc{ii} regions and analyze their single-slit spectra in the light of these models to understand their structure and emission properties. Our models are one-dimensional (1D), and hence we tried to fit to the closest thing to 1D models: slit spectra of spherical homogeneous regions. The purpose of this paper is to introduce a new framework and motivate its usage, and we are already planning to expand the work to 3D models on multiple LVM objects.

Our models performed well in duplicating key diagnostic line fluxes for the three H\,\textsc{ii} regions studied, and the inferred physical conditions seem to agree with the evidence we have from observations. Nevertheless, the search method we applied to find the best-fit model is rather simple, and there is a degeneracy of models that would satisfy the minimum $\chi^2$ criterion. Further constraints and an advanced search process. 

One option is to consider the spatially resolved spectra of these regions rather than looking at their aperture-integrated spectra from the brightest regions. The idea is to constrain the temperature and density diagnostic line ratios in each spatial bin, as well as their average integrated value, to better study the internal ionization structure and more accurately model the nebula. We present one such case here, where we consider both single-slit and spatially resolved observations of the H\,\textsc{ii} region M\,20. This experiment allowed us to better select the best-fit model, and as a result, the model better represents the internal temperature and density profile in addition to the key line fluxes. Moreover, the density estimates using either observation give us different results. This could be due to observational issues or the fact that some of the important lines are detected in the single slit but not with the spatially resolved data. It is interesting that the best-fit model turned out to be approximately uniform for this region. Alternatively, it could be that the inhomogeneities in M\,20 are below the spatial resolution of the LVM data, and that the variations are so small-scale that both the observations and the model lack the capabilities to resolve them.

Moreover, most H\,\textsc{ii} regions are not spherically symmetric like M\,20 and probably have multiple sources of ionizing photons, such as shocks, protostars, and external sources. In that case, the variations in the physical conditions would not only be across the plane of the sky but also along the line of sight. Since our models are 1D, for more complicated systems it would be better to model each spatial pixel with a 1D model and do the same for a group of neighboring pixels to try to understand the nature of the variations across different pixels. We plan to do this with M42 (Orion), as it is one of the brightest sources in the sky, is confirmed to have a blister-type configuration, and is being observed by the LVM \citep{Kreckel+2024}. Furthermore, we have only relied on the optical diagnostic lines to predict abundances, but this is usually better done with multi-wavelength observations since some species (e.g., C$^{+2}$ and O$^{+3}$) don't have lines in the optical and are necessary to establish the ionization balance and predict ionic fractions.

In conclusion, our findings demonstrate that accounting for variations in temperature, density, and chemical profiles when modeling the emission spectra of photoionized nebulae is capable of capturing their complex, inhomogeneous internal structures and inferring their true chemical composition. This aligns with the solution proposed by \citet{Peimbert+1967} to address the AD problem, a hypothesis that has been supported by both observational \citep{Mendez+2022,Jin+2023} and theoretical \citep{Marconi+2024,Tsamis+2005} studies. Although the presence of temperature and chemical fluctuations in photoionized nebulae is widely acknowledged in the literature, there has been limited theoretical work to quantify the impact of these fluctuations. A few studies, utilizing multi-cloud simulations where the emission is integrated across different components \citep{Marconi+2024,Tsamis+2005}, have made important progress toward inhomogeneous theoretical models. Our models perform comparably to previous models but provide a novel framework and demonstrate its feasibility to investigate photoionized nebulae under the assumption of inhomogeneity. However, based on the limited sample of objects we have compared, it remains unclear what proportion of H\,\textsc{ii} regions exhibit chemical inhomogeneities (e.g., M\,17), temperature fluctuations (e.g., NGC 2579), or are more uniform and can be approximately modeled as homogeneous systems, as in the case of M\,20.

Future work will extend this analysis to a broader sample of H,\textsc{ii} regions, supported by existing and forthcoming integral field unit (IFU) observations and multi-wavelength data. We are already working to expand the work to 3D models on multiple LVM objects. This will allow us to address some of the limitations of the current 1D models, such as treating slit spectra as if they are 1D when they are not, and improve the comparison among different families for better or worse fits. By advancing the theoretical modeling of photoionized nebulae, we aim to better understand their complex structures and ultimately resolve the abundance discrepancy problem.

\newpage

 \begin{table}[!h]
%    \centering
    \hspace*{-0.4in}
    \begin{tabular}[b]{cccc}\hline
      line & observed & $\Delta_f$\% &$\Delta_h$\%\\ \hline
      $[O\,\textsc{ii}]$ 3726 &  146.6  & 2.73 & 51.5\\
      $[O\,\textsc{ii}]$ 3729 &  173.3  & 2.68 & 47.0\\
      $[S\,\textsc{ii}]$ 4069 &  1.802  & 24.8 & 90.3\\
      $[S\,\textsc{ii}]$ 4076 &  0.627  & 32.8 & 90.2\\
       C\,\textsc{ii}  4267   &  0.17   & 34.9 & 64.4\\
      $[O\,\textsc{iii}]$ 4363 &  0.15   & 15.5& 58.3\\
       O\,\textsc{ii}  4641   &  0.03   & 26.2 & 74.4\\
      $[O\,\textsc{iii}]$ 4959 &  19.0   & 38.7& 74.5\\
      $[O\,\textsc{iii}]$ 5007 &  58.9   & 35.0& 9.8\\
      $[N\,\textsc{ii}]$ 5755 &  0.97   & 8.75 & 36.9\\
      $[N\,\textsc{ii}]$ 6584 &  110.9  & 26.4 & 119.5\\
      $H\alpha$               &  286.7  & 0.22 & 52.8\\
      He\,\textsc{i} 6678     &  2.983   & 13.1 & 52.8\\
      $[S\,\textsc{ii}]$ 6716 &  24.1   & 21.1 & 81.5\\
      $[S\,\textsc{ii}]$ 6730 &  21.29  & 23.1 & 83.7\\
      He\,\textsc{i} 7281     &  0.522   & 17.6 & 54.6\\
      $[O\,\textsc{ii}]$ 7319 &  0.543  & 16.0 & 74.8\\
%      $[O\,\textsc{ii}]$ 7320 &  1.876  & 55.5 \\
%      $[O\,\textsc{ii}]$ 7330 &  1.339  & 55.5 \\
%      $[O\,\textsc{ii}]$ 7331 &  1.004  & 55.5 \\
     \hline
      parameter &  & value \\ \hline
      log $T_{eff}(K)$   &     &  4.60 &    \\
      log $L^*(erg/s)$       &     &  38.6 &    \\
      $scale_{grain}$       &     &  4.6 &    \\
     \hline
      density sinusoidal &  & value \\ \hline
      log(Period(cm))       &     &  19.3 &    \\
      max($cm^{-3}$)       &     &  263 &    \\
      min($cm^{-3}$)       &     &  232 &    \\
      phase(rad)       &     &  0.6 &    \\
     \hline
      abundance sinusoidal &  & value \\ \hline
      log(Period(cm))       &     &  19.2 &    \\
      max       &     &  1.59 &    \\
      min       &     &  0.83 &    \\
      phase(rad)       &     &  0 &    \\
     \hline
      quantity & observed & prediction \\ \hline
      $n_e (cm^{-3})$    &       &     &  \\
      $[O\,\textsc{ii}]$ 3726/3729 &  $240\pm70$  & 236 &  \\
      $[S\,\textsc{ii}]$ 6716/6730 &  $320\pm130$  & 343 &  \\
%      $n_e (adopted)$   &  $270\pm60$  & 184 \\
      $T_e (K)$  &        &     &  \\
      $[N\,\textsc{ii}]$ 6584/5755 &  $8500\pm240$  & 8857 &  \\
      $[O\,\textsc{iii}]$ 4959/4363 &  $7800\pm300$  & 7409 &  \\    
    \end{tabular}
    \qquad
    \begin{tabular}[b]{cccc}\hline
      line & observed & $\Delta_f$\% &$\Delta_h$\%\\ \hline
      $[O\,\textsc{ii}]$ 3726 &  146.6  & 5.58 & 54.5\\
      $[O\,\textsc{ii}]$ 3729 &  173.3  & 9.85 & 50.6\\
      $[S\,\textsc{ii}]$ 4069 &  1.802  & 56.1 & 91.3\\
      $[S\,\textsc{ii}]$ 4076 &  0.627  & 63.4 & 91.2\\
       C\,\textsc{ii}  4267   &  0.17   & 27.6 & 59.0\\
      $[O\,\textsc{iii}]$ 4363&  0.15   & 84.8 & 53.3\\
       O\,\textsc{ii}  4641   &  0.03   & 58.3 & 69.5\\
      $[O\,\textsc{iii}]$ 4959&  19.0   & 82.1 & 69.5\\
      $[O\,\textsc{iii}]$ 5007&  58.9   & 78.9 & 9.5\\
      $[N\,\textsc{ii}]$ 5755 &  0.97   & 2.85 & 37.2\\
      $[N\,\textsc{ii}]$ 6584 &  110.9  & 18.9 & 116.8\\
      $H{\alpha}$             &  286.7  & 0.20 & 53.1\\
      He\,\textsc{i} 6678     &  2.983   & 13.4 & 53.1\\
      $[S\,\textsc{ii}]$ 6716 &  24.1   & 46.0 & 76.6\\
      $[S\,\textsc{ii}]$ 6730 &  21.29  & 51.2 & 83.5\\
      He\,\textsc{i} 7281     &  0.522   & 0.95 & 49.6\\
      $[O\,\textsc{ii}]$ 7319 &  0.543  & 22.0 & 70.4\\
%      $[O\,\textsc{ii}]$ 7320 &  1.876  & 55.5 \\
%      $[O\,\textsc{ii}]$ 7330 &  1.339  & 55.5 \\
%      $[O\,\textsc{ii}]$ 7331 &  1.004  & 55.5 \\
     \hline
      parameter &  & value \\ \hline
      log $T_{eff}(K)$   &     &  4.50 &    \\
      log $L^*(erg/s)$       &     &  38.9 &    \\
      $scale_{grain}$       &     &  1.1 &    \\
     \hline
      density sinusoidal &  & value \\ \hline
      log(Period(cm))       &     &  19.4 &    \\
      max($cm^{-3}$)       &     &  219 &    \\
      min($cm^{-3}$)       &     &  155 &    \\
      phase(rad)       &     &  0 &    \\
     \hline
      abundance sinusoidal &  & value \\ \hline
      log(Period(cm))       &     &  19.4 &    \\
      max       &     &  1.33&    \\
      min       &     &  0.83 &    \\
      phase(rad)       &     &  0.1 &    \\
     \hline
      quantity & observed & prediction \\ \hline
      $n_e (cm^{-3})$    &       &     & \\
      $[O\,\textsc{ii}]$ 3726/3729&  $240\pm70$  & 182 & \\
      $[S\,\textsc{ii}]$ 6716/6730&  $320\pm130$  & 270 & \\
%      $n_e (adopted)$   &  $270\pm60$  & 184 \\
      $T_e (K)$  &        &     &  \\
      $[N\,\textsc{ii}]$ 6584/5755&  $8500\pm240$  & 8857 & \\
      $[O\,\textsc{iii}]$ 4959/4363&  $7800\pm300$  & 7409 & \\
    \end{tabular}
    \caption{Left: Best fit model for the single slit observation of M\,20 belonging to family (3) where we introduce temperature and density inhomogeneities. Right: Best fit model for the LVM observation of M\,20 belonging to family (3). We show in these tables the percent difference between our predictions and the observation for key CELs and RLs normalized such that $H{\beta}=100$.}
  \label{tab:1}
\end{table}
\newpage

 \begin{table}[!ht]
%    \centering
     \hspace*{-0.5in}
    \begin{tabular}[b]{cccc}\hline
      line & observed & $\Delta_f$\% &$\Delta_h$\%\\ \hline
      $[O\,\textsc{ii}]$ 3726 &  45.6  & 11.6 & 22.9\\
      $[O\,\textsc{ii}]$ 3729 &  46.2  & 28.3 & 24.6\\
      $[S\,\textsc{ii}]$ 4069 &  0.427  & 13.8& 39.0\\
      $[S\,\textsc{ii}]$ 4076 &  0.131  & 9.25& 38.7\\
       C\,\textsc{ii}  4267   &  0.58   & 62.5& 113.8\\
      $[O\,\textsc{iii}]$ 4363 & 0.95   & 74.6& 164.1\\
       O\,\textsc{ii}  4641   &  0.12   & 44.6& 145.5\\
      $[O\,\textsc{iii}]$ 4959 & 114.1  & 35.0& 108.0\\
      $[O\,\textsc{iii}]$ 5007 & 335.4  & 33.5& 108.0\\
      $[N\,\textsc{ii}]$ 5755 &  0.26   & 30.4& 62.2\\
      $[N\,\textsc{ii}]$ 6584 &  24.5  & 5.49 & 31.8\\
      $H\alpha$               & 285.3  & 3.09 & 3.8\\
      He\,\textsc{i} 6678     &  3.946  & 1.31 & 10.4\\
      $[S\,\textsc{ii}]$ 6716 &  4.54   & 29.6& 36.7\\
      $[S\,\textsc{ii}]$ 6730 &  4.35  & 18.0 & 34.9\\
      He\,\textsc{i} 7281     &  0.589  & 5.59 & 13.9\\
      $[O\,\textsc{ii}]$ 7319 &  0.177  & 12.0& 68.8\\
%      $[O\,\textsc{ii}]$ 7320 &  1.876  & 55.5 \\
%      $[O\,\textsc{ii}]$ 7330 &  1.339  & 55.5 \\
%      $[O\,\textsc{ii}]$ 7331 &  1.004  & 55.5 \\
      \hline
      parameter &  & value \\ \hline
      log $T_{eff}(K)$   &     &  4.59   \\
      log $L^*(erg/s)$       &     &  41.0   \\
      $scale_{grain}$       &     &  2.4 &    \\
     \hline
      abundance sinusoidal &  & value \\ \hline
      log(Period(cm))       &     &  19.7 &    \\
      max       &     &  1.7 &    \\
      min       &     &  0.6 &    \\
      phase(rad)       &     &  0.25 &    \\
     \hline
      quantity & observed & prediction \\ \hline
      $n_e (cm^{-3})$    &       &     \\
      $[O\,\textsc{ii}]$ 3726/3729 &  $480\pm150$  & 210 \\
      $[S\,\textsc{ii}]$ 6716/6730 &  $500\pm220$  & 307 \\
%      $n_e (adopted)$   &  $270\pm60$  & 184 \\
      $T_e (K)$  &        &     \\
      $[N\,\textsc{ii}]$ 6584/5755 &  $8950\pm380$  & 8350 \\
      $[O\,\textsc{iii}]$ 4959/4363 &  $8020\pm170$  & 7207 \\    
    \end{tabular}
    \qquad
    \begin{tabular}[b]{cccc}\hline
      line & observed & $\Delta_f$\% &$\Delta_h$\%\\ \hline
      $[O\,\textsc{ii}]$ 3726 & 64.9  & 34.7 & 66.3\\
      $[O\,\textsc{ii}]$ 3729 & 50.6  & 41.8 & 64.8\\
      $[S\,\textsc{ii}]$ 4069 &  0.67  & 39.8& 79.9\\
      $[S\,\textsc{ii}]$ 4076 &  0.209 & 37.6& 80.0\\
       C\,\textsc{ii}  4267   & 0.16  & 65.0 & 121.9\\
      $[O\,\textsc{iii}]$ 4363& 1.88  & 0.54 & 117.8\\
       O\,\textsc{ii}  4641   & 0.07  & 50.9 & 119.6\\
      $[O\,\textsc{iii}]$ 4959& 122.8 & 14.0 & 119.6\\
      $[O\,\textsc{iii}]$ 5007& 356.2 & 24.6 & 40.5\\
      $[N\,\textsc{ii}]$ 5755 & --    & --   & 120.1\\
      $[N\,\textsc{ii}]$ 6584 & 23.8  & 1.9  & 119.1\\
      $H{\alpha}$             &  290.0 & 12.0& 89.3\\
      He\,\textsc{i} 6678     &  3.508 & 3.88 & 91.9\\
      $[S\,\textsc{ii}]$ 6716 & 3.48 & 3.02  & 92.2\\
      $[S\,\textsc{ii}]$ 6730 & 4.10  & 23.8 & 77.1\\
      He\,\textsc{i} 7281     & 0.671 & 21.1  & 44.4\\
      $[O\,\textsc{ii}]$ 7319 & 0.765 & 25.5 & 44.5\\
%      $[O\,\textsc{ii}]$ 7320 &  1.876  & 55.5 \\
%      $[O\,\textsc{ii}]$ 7330 &  1.339  & 55.5 \\
%      $[O\,\textsc{ii}]$ 7331 &  1.004  & 55.5 \\
      \hline
      parameter &  & value \\ \hline
      log $T_{eff}(K)$   &     &  3.98   \\
      log $L^*(erg/s)$       &     &  41.3   \\
      $scale_{grain}$       &     &  4.0 &    \\
     \hline
      abundance sinusoidal &  & value \\ \hline
      log(Period(cm))       &     &  19.2 &    \\
      max       &     &  1.47 &    \\
      min       &     &  0.915 &    \\
      phase(rad)       &     &  0.35 &    \\
     \hline
      quantity & observed & prediction \\ \hline
      $n_e (cm^{-3})$    &       &     \\
      $[O\,\textsc{ii}]$ 3726/3729&  $1570\pm280$  & 849 \\
      $[S\,\textsc{ii}]$ 6716/6730&  $1140\pm230$  & 1274 \\
%      $n_e (adopted)$   &  $270\pm60$  & 184 \\
      $T_e (K)$  &        &     \\
      $[N\,\textsc{ii}]$ 6584/5755 &  --  & -- \\
      $[O\,\textsc{iii}]$ 4959/4363 &  $9410\pm160$  & 8297 \\
       
    \end{tabular}
    \caption{Left: Best fit model for the single slit observation
of M\,17 belonging to family (1) where we introduce chemical inhomogeneities. Right: Best fit model for the single slit observation
of NGC\,2579 belonging to family (2) where we introduce temperature inhomogeneities. We show in these tables the percent difference between our predictions and the observation for key CELs and RLs normalized such that $H{\beta}=100$.}
  \label{tab:2}
\end{table}

\section*{Acknowledgments}
JEMD, NS and KK acknowledge support from the Deutsche Forschungsgemeinschaft (DFG, German Research Foundation) in the form of an Emmy Noether Research Group (grant number KR4598/2-1, PI Kreckel) and the European Research Council’s starting grant ERC StG-101077573 (“ISM-METALS"). Funding for the Sloan Digital Sky Survey V has been provided by the Alfred P. Sloan Foundation, the Heising-Simons Foundation, the National Science Foundation, and the Participating Institutions. SDSS acknowledges support and resources from the Center for High-Performance Computing at the University of Utah. SDSS telescopes are located at Apache Point Observatory, funded by the Astrophysical Research Consortium and operated by New Mexico State University, and at Las Campanas Observatory, operated by the Carnegie Institution for Science. The SDSS web site is \url{www.sdss.org}. G.A.B. acknowledges the support from the ANID Basal project FB210003.

SDSS is managed by the Astrophysical Research Consortium for the Participating Institutions of the SDSS Collaboration, including the Carnegie Institution for Science, Chilean National Time Allocation Committee (CNTAC) ratified researchers, Caltech, the Gotham Participation Group, Harvard University, Heidelberg University, The Flatiron Institute, The Johns Hopkins University, L'Ecole polytechnique f\'{e}d\'{e}rale de Lausanne (EPFL), Leibniz-Institut f\"{u}r Astrophysik Potsdam (AIP), Max-Planck-Institut f\"{u}r Astronomie (MPIA Heidelberg), Max-Planck-Institut f\"{u}r Extraterrestrische Physik (MPE), Nanjing University, National Astronomical Observatories of China (NAOC), New Mexico State University, The Ohio State University, Pennsylvania State University, Smithsonian Astrophysical Observatory, Space Telescope Science Institute (STScI), the Stellar Astrophysics Participation Group, Universidad Nacional Aut\'{o}noma de M\'{e}xico, University of Arizona, University of Colorado Boulder, University of Illinois at Urbana-Champaign, University of Toronto, University of Utah, University of Virginia, Yale University, and Yunnan University.

\end{CJK*}
\end{document}